%% file: main.tex
\documentclass[12pt]{article}
\usepackage{fullpage}
\usepackage[margin=1in]{geometry}
\usepackage[utf8]{inputenc}

\title{The Rational Agent Benchmark for Data Visualization}

\author{Yifan Wu \footnote{Department of Computer Science, Northwestern University.
Email: \texttt{yifan.wu@u.northwestern.edu}.}
\and Ziyang Guo \footnote{Department of Computer Science, Northwestern University.
Email: \texttt{ziyangguo2027@u.northwestern.edu}.}
\and Michalis Mamakos\footnote{Department of Computer Science, Northwestern University.\\
 Email: \texttt{michailmamakos2022@u.northwestern.edu}}
\and Jason Hartline\footnote{Department of Computer Science, Northwestern University.
Email: \texttt{hartline@northwestern.edu}}
\and Jessica Hullman\footnote{Department of Computer Science, Northwestern University.
Email: \texttt{jhullman@northwestern.edu}}
}

\usepackage{tabu}                      
\usepackage{booktabs}                  
\usepackage{lipsum}                    
\usepackage{mwe}                       

\usepackage{mathptmx}                  

\usepackage[svgnames]{xcolor}

\usepackage[makeroom]{cancel}
\usepackage{amsmath}
\usepackage{amssymb}
\usepackage{amsthm}
\usepackage{amsfonts}
\usepackage{mathtools}
\usepackage{bbm}
\usepackage{bm}
\usepackage{verbatim}
\usepackage{booktabs}
\usepackage{array}
\usepackage{setspace}
\usepackage{caption}
\usepackage{subcaption}
\usepackage[export]{adjustbox}
\usepackage{float}
\usepackage{changepage}
\usepackage{enumitem}
\usepackage{cases}
\usepackage{soul}
\usepackage{hyperref}
\usepackage{longtable}
\usepackage{algorithm}
\usepackage{algpseudocode}
\usepackage{cleveref}
\usepackage{multirow}

\definecolor{calibratedcolor}{RGB}{217, 95, 2}
\definecolor{behavioralcolor}{RGB}{117,112,179}
\definecolor{behavioralcolordecision}{RGB}{27,158,119}

\newcommand{\calibratedtextcolor}[1]{\textcolor{calibratedcolor}{#1}}
\newcommand{\behavioraltextcolor}[1]{\textcolor{behavioralcolor}{#1}}
\newcommand{\behavioralactiontextcolor}[1]{\textcolor{behavioralcolordecision}{#1}}

\DeclareMathOperator*{\argmax}{arg\,max}

\DeclareMathOperator{\E}{\mathbb{E}}

\input{notation}

\usepackage{natbib}

\begin{document}




\maketitle
\abstract{%
Understanding how helpful a visualization is from experimental results is difficult because the observed performance is confounded with aspects of the study design, such as how useful the information that is visualized is for the task. We develop a rational agent framework for designing and interpreting visualization experiments. Our framework conceives two experiments with the same setup: one with behavioral agents (human subjects), and the other one with a hypothetical rational agent. A visualization is evaluated by comparing the expected performance of behavioral agents to that of a rational agent under different assumptions. Using recent visualization decision studies from the literature, we demonstrate how the framework can be used to pre-experimentally evaluate the experiment design by bounding the expected improvement in performance from having access to visualizations, and post-experimentally to deconfound errors of information extraction from errors of optimization, among other analyses. 
}


\let\cite=\citep

\section{Introduction}
\input{00_intro}

\input{01_related}

\input{02_framework}

\input{03_demonstrations}
\input{04_discussion}

\input{05_ack}


\bibliographystyle{unsrtnat}
\bibliography{ref}

\end{document}

%% file: notation.tex
\newcommand{\rperfect}{\textsc{R}_V^R}
\newcommand{\rpos}{\textsc{R}_V}
\newcommand{\rprior}{\textsc{R}_{\varnothing}}
\let\rscore=\rpos

\newcommand{\bactionraw}{\textsc{B}}
\newcommand{\bactioncalib}{\textsc{R}_B}
\newcommand{\bresponseraw}{B^Q}
\newcommand{\bresponsecalib}{C^Q}
\newcommand{\responsespace}{Q}

\let\bscore=\bactionraw
\let\cscore=\bactioncalib

\newcommand{\state}{\theta}
\newcommand{\statespace}{\Theta}

\newcommand{\finestate}{x}
\newcommand{\finestatespace}{X}

\newcommand{\finestatemap}{\hat{\state}}

\newcommand{\joint}{\pi}
\newcommand{\bjoint}{\joint^B}

\newcommand{\dist}{p}
\newcommand{\prior}{p}
\newcommand{\posterior}{q}

\newcommand{\distover}[1]{\Delta(#1)}
\newcommand{\indicator}[1]{\mathbf{1}[#1]}

\newcommand{\action}{a}
\newcommand{\actionspace}{A}

\newcommand{\signal}{v}
\newcommand{\signalspace}{V}

\newcommand{\score}{S}
\newcommand{\proper}{\hat{\score}}

\newcommand{\reals}{{\mathbb R}}

\newcommand{\infoval}{\Delta}
\newcommand{\binfoval}{\infoval^B}

\newcommand{\reward}{r}

\newcommand{\maxtime}{T}

\newcommand{\convrate}{d}

\usepackage{ifthen}

\newcommand{\prob}[2][]{\text{\bf Pr}\ifthenelse{\not\equal{}{#1}}{_{#1}}{}\![{\def\givenn{\middle|}#2}]}
\newcommand{\expect}[2][]{\text{\bf E}\ifthenelse{\not\equal{}{#1}}{_{#1}}{}\![{\def\givenn{\middle|}#2}]}

\newcommand{\tabitem}{~~\llap{\textbullet}~~}

%% file: 00_intro.tex
\label{sec:intro}
Intuition-driven design guidelines for designing data visualizations are increasingly being replaced with data-driven recommendations based on visualization studies.  
To assess the extent to which modern empirical study of visualizations does in fact capture the value of visualization, however, requires accounting for the design of the experimental task and conditions for study. To understand how well people performed with a visualization in a controlled study, or how important an observed difference in performance between two visualization is, we must understand what sorts of performance differences an experimental scenario admits. 

However, it can be difficult in designing a study to predict how the choices one makes impact the experiment's capability for capturing meaningful performance differences. We can liken the experiment design process to setting various "knobs" that will impact the difficulty of the task, the extent to which participants are motivated to study the visualization to complete the task, and the best achievable performance on the task. These knobs include the input distributions used to generate stimuli, the allocation of these inputs across participants, and the payoff function that will reward participants for making good decisions. More broadly applicable experiment design decisions include how many participants to target and how to compare key interventions (e.g., between-subjects, pre-post design, etc.). 

While it is difficult to define "optimal" choices for these myriad decisions, the results of a study can still provide useful knowledge about visualization performance when properly conditioned on the \textit{potential} the study had, whether to show differences between visualization strategies or to evaluate a specific strategy. For example, a canonical form of conditioning used to assess a study's potential to detect an effect such as a difference between treatments ensures that the study design provides sufficient statistical power to detect an effect of the hypothesized size. 

More generally, we would like an approach to interpreting the results of a study comparing visualization strategies that helps a reader answer questions like the following:

\begin{itemize}[itemsep=1pt, leftmargin=9pt]
\item How hard is the task? For example, how well could we expect someone do without consulting the visualized data at all? 
\item Considering the study design alone, how incentivized would we expect participants to be to use the visualized information?
\item To what extent are observed differences in performance likely to stem from informational asymmetries in the visualizations (e.g., visualizing only a mean versus a more expressive depiction of a distribution)?
\item To what extent is sub-optimal performance with a visualization due to participants not differentiating the task-relevant information it provides, versus not being able to properly use the information they gained to choose a response?
  
\end{itemize}


Our inability to answer the above questions from many empirical research papers highlight how visualization research lacks clear comparison points, or performance \textit{benchmarks} that can aid the design and interpretation of controlled visualization experiments.
Answering such questions contextualizes what was learned from observing the performance of any single visualization in absolute terms defined on the experiment design.
 Without clear benchmarks, readers and authors alike tend to draw conclusions from coarse, \textit{relative} information like visualization performance rankings. 
 A good set of benchmarks are necessary to assess the fitness of the experiment design itself for studying a given visualization research question.

We contribute a rational agent framework based on quantifying the value of information to a judgment or decision problem. Our framework defines benchmark measures representing attainable performance given a visualization experiment design. 
Benchmarks defined in the rational agent framework can be applied before an experiment is run to vet how capable the experiment design is of showing important differences between visualizations and of resolving good performance with any single visualization.
Applying the framework after an experiment provides further insight into behavioral agent performance, by enabling the researcher to deconfound sources of erroneous answers. For example, agents might be unable to extract the information from the visualization, or unable to optimally translate the information to a decision. 

We apply the framework to two well-regarded visualization experiments from the literature: one on the impact of visualization design on effect size judgments and decisions~\cite{kale2020visual} and one on the impact of visualization design on transit decisions~\cite{fernandes2018uncertainty}. In both cases, we identify 1) ways in which the experiment design could have been improved (through different measures or payoff functions) and 2) sources of loss that help explain behavioral results but were not fully addressed in the original presentations of results. 

%% file: 01_related.tex
\section{Related Work}

\subsection{Visualization Evaluation}


Our work aims to improve evaluation methods in visualization. Previously, researchers have contributed overviews of qualitative and quantitative approaches~\cite{isenberg2013,lam2012,zuk2006} and conceptual models and approaches for ensuring that one selects an evaluation that is appropriate for a given task, context, or contribution type~\cite{isenberg2008,munzner,shneiderman2006}.


 
Whenever visualizations are meant to support inference in addition to merely describing an observed dataset~\cite{hullman2021designing},
the evaluation approach should define a standard for assessing the quality of the inference.
However, several recent surveys of evaluative studies for visualizations~\cite{dimara2021critical} and uncertainty visualizations specifically~\cite{hullman2018pursuit,kinkeldey2014} suggest that the use of well-defined judgment and decision tasks is rare. Instead, a majority of uncertainty visualization studies rely on measures of perceptual accuracy and/or self-reports of satisfaction, confidence, or other properties that may have an unclear or even opposite relationship with rational use of the information for the problem at hand~\cite{hullman2018pursuit,kinkeldey2014}.
This has led some researchers to advocate for adopting 
Bayesian inference as a benchmark against which to compare reactions to visualizations~\cite{hullman2021designing,kale2021causal,kim2020bayesian}. These models use the deviation of human performance from the Bayesian ideal as a means of better understanding patterns in human judgments, and for inspiring new design approaches~\cite{hullman2021designing,kale2023evm}. 
While human judgments need not be perfectly Bayesian for such approaches to lead to a better understanding of how people use visualizations, if there is no correspondence between human behavior and the Bayesian agent's behavior, design suggestions aimed at aligning the human behavior with the Bayesian's them may not be effective. 
In contrast to prior applications of Bayesian theory to visualization, the value of the rational agent framework does not depend on actual humans acting like rational agents.  
Our work is related to ideal observer analysis, used in psychophysics, which theoretically upperbounds behavioral performance by a Bayesian agent in the same situation in order to reason about factors influencing human perception \cite{KW-96}. However, our framework defines the baseline performance in addition to the upperbound, and hence provides a ``scale'' for interpreting behavioral performance and a means to separate sources of loss in decision-making. 

\subsection{Interpreting experiment results}
Our work is related to recent integrative modeling~\cite{hofman2021integrating} approaches to benchmarking the irreducible variance in data used for modeling~\cite{agrawal2020scaling,Fudenberg2022MeasuringTC}.
For example, the explanatory power of theories embedded in behavioral models can be assessed by quantifying irreducible error inherent in an experimental task~\cite{Fudenberg2022MeasuringTC},  grounding a perspective for how well a model performs. We take a similar approach, but with the goal of benchmarking how well humans can be expected to do under different assumptions when faced with an experimental task. 

%% file: 02_framework.tex
\section{The Rational Agent Framework}
\label{sec:framework} 
The value of the information presented in a visualization can be quantified by how much it improves the expected payoff in a decision problem.  
The visualized information reduces uncertainty about a payoff-relevant state, thus helping the agent make better decisions. The value of the visualization can be understood as the expected improvement in payoff when an agent has access to the visualization. 

Our framework conceives two studies, an experimental study and a theoretical one. The first occurs in the real world with behavioral participants, and the other is based on an analysis of a hypothetical rational world with a rational agent participant. 
We assume an experiment design as input, including information on how stimuli will be generated, what decisions or beliefs participants will report, and how their responses will be incentivized and scored. 
If the experiment has already been conducted, the raw or modeled behavioral results are also part of the input.
The two studies assume exactly the same decision problem and data-generating process, enabling analysis of an experiment both before and after it is run.

Below we establish preliminaries, including what constitutes a visualization experiment in our framework, the conceptual devices of the rational and behavioral agent, and how they are used in pre- and post-experimental analyses. We apply these definitions to an example forecast visualization experiment. 




\input{02a-basic-framework}

\subsection{Applying the Framework to Visualization Studies}

\subsubsection{Scope: What is a decision experiment?}

The rational agent framework can be applied widely across empirical visualization studies. To apply the framework the experiment task needs to involve the visualization of states that can take on multiple values and under which the rational agent's optimal decision -- for payoff or accuracy -- is non-identical.  In such experiments, the rational benchmark and the rational baseline are distinct and there is a non-trivial value of information.

It is worth noting that our use of the term “decision” aligns with statistical decision theory, and may conflict with colloquial interpretations promoted elsewhere in visualization research. For example, we could apply the framework to perception studies (like Cleveland and McGill's well-known position-length experiment~\cite{cleveland1984graphical}) and refer to the task participants face as a decision task. The uncertainty in the state comes from the fact that there is a distribution over ground truth proportions that are used to generate stimuli.

There are just two conditions that prevent applying the rational agent framework.  The first is in studies where there is no differing state.  For example, if the exact same data are presented to all participants in a single-trial between-subjects manipulation of visualization design then there is no uncertainty about the state and the rational benchmark and baseline would coincide.  The second is in studies for which the experimenter considers it impossible to define a ground-truth response against which to evaluate participants' reports, such as studies that query agents' emotional states (e.g., angry, excited, sad) after showing a visualization.  For such studies, optimal reports by a rational agent are not well defined.

In decision experiments, scoring rules are typically used to incentivize the behavioral agent to make good decisions and to evaluate the quality of the decision made, such as the accuracy of a prediction.  
The experimenter may use the same scoring rule for both incentives and accuracy; or the experimenter may not incentivize the behavioral agent at all. For example, it is not clear if participants in the position-length experiment~\cite{cleveland1984graphical} were compensated more for doing the tasks well, but mid-mean absolute error is used to evaluate their responses.  The rational agent framework applied to either scoring rules for incentives or accuracy can help understand how effectively information is conveyed by a visualization; the framework's application to scoring rules for incentives can additionally help understand the potential effectiveness of the incentives.

For any decision task, we can distinguish between the decision--the reported ``action''--and the beliefs that led to that decision. 
However, when a decision is defined on a coarse action space, such as binary, calibration will be of limited use, because multiple different beliefs will lead to the same decision so the decision is not informative about the agent's belief. Recall that the optimization loss is the difference $\bactioncalib - \bactionraw$ between the calibrated score and the raw score. When the calibrated score is not informative about the agent's optimal payoff as dictated by belief, the experimenter does not estimate the optimization loss precisely.  
Hence, an experimenter could potentially better quantify the usefulness of the visualization by refining the action space or asking for beliefs directly, i.e., with the action space $\actionspace = \distover{\statespace}$, the set of distributions over states. 



\subsubsection{$\rprior$ as a simple baseline}
The rational baseline $\rprior$ captures what a rational agent would do in the experiment if they didn't look at the visualizations. 
This concept is novel in visualization research, where attempts to detect reliance on visualizations remain relatively rare. Instead, observed performance is usually compared only to the best possible performance for the task, as in computing perceptual or decision accuracy. 

We can compare $\rprior$ to different notions of a simple baseline that an experimenter might use to simulate a behavioral agent not paying attention. For example, a researcher might consider random response over the allowable values for the measure (e.g., randomly choosing a value between 0 and 100 for a task that elicits an integer-valued probability) as a useful simple baseline, or designing a study specifically to compare observed behavior to expectations under a heuristic (e.g. Kale et al.\cite{kale2020visual}).
There is nothing wrong with using other simple baselines to estimate bad performance. However, the unique value of $\rprior$ as a definitive benchmark is for separating cases where participants got information from the visualization from cases where they did not. 
If we use other forms of ``random guessing'' as the baseline, agents could still not look at the visualization at all and do better than the random baseline, so long as random guessing performs worse in expectation than using the prior. Only observing that agents did better than the prior lets us evaluate a “null hypothesis” that they did not consult the visualization.

The fact that the prior is not provided to participants in many visualization experiments does not affect its value for evaluating the state of evidence on whether agents consulted the visualization. In some cases, even when a prior is not provided, $\rprior$ may still be a realistic expectation of how participants who are not carefully consulting the visualization would respond. 
For example, when the experiment involves repeated measures (trials) and agents receive feedback, with enough trials we might expect behavioral agents to achieve the expected payoff $\rprior$ by learning that some fixed action guarantees an okay payoff without looking at the visualization. Research into learning from samples (e.g.\ Gonzalez and Dutt\cite{gonzalez2011instance}) can inform speculation about particular repeated feedback experiment designs.

\subsubsection{Calculating behavioral scores}

$\rpos$, the rational agent's payoff under the action dictated by their posterior beliefs, represents the best attainable performance by a behavioral agent who does the experiment.  Whenever the goal of the experiment is to compare the performance of visualization strategies that differ in the information they provide for the task, $\rpos$ and $\Delta$ can be calculated for each visualization condition tested. Different visualization optimal $\rpos$ for 
informationally-inequivalent visualizations give us a sense of how much the results of the experiment can be driven purely by information differences. 
In general, researchers who are interested in understanding differences that result from visual design choices, rather than informational differences, should aim for equivalent visualization optimal $\rpos$. 
Exceptions include cases where the goal is to investigate how visualization approaches compare for a real-world inspired task where a conventional representation may not be richly informative, such as situations where point estimates are preferred by convention~\cite{hullman2019authors}. 
Whenever informationally-inequivalent visualizations are compared, the experimenter can use the information loss $(\rperfect-\rpos)/\infoval$ to study 
the maximum differences we expect under optimal use of the two visualizations. 
\footnote{Additionally, we can use comparisons between information loss for informationally-different visualizations to weed out claims a researcher makes about one visualization being informationally superior than another: A larger effect than the difference in the two $\rpos$ that is claimed to result from informationally-inequality must be an overestimate. More generally, any experiment that presents estimates corresponding to a higher expected score under the scoring rule for a given visualization must be presenting an overestimate confounded, for example, by sampling  error~\cite{button2013power}.}

Generally, we employ estimates of joint behavior of the agent with the state, $\bjoint \in \distover{\actionspace \times \statespace}$, from a statistical model that accounts for the design of the experiment. This is because rarely can the results of an experiment be interpreted without accounting for confounding induced by the design in the form of order effects, random effects of participants or other factors, etc. 
The target in producing model estimates of $\bjoint$ is to achieve a good prediction of the score distribution expected for behavioral agents if the experiment were to be repeated many times on a new sample from the same population.
In general, \textit{generative} statistical models that model the joint probability distribution $p(x,y)$ and use Bayes rule to compute $p(y|x)$ are preferable.
For example, in our demonstrations below, we use Bayesian regression models. However, our approach is compatible with sampling from observed results directly or using non-generative models (e.g., Frequentist regression), as long as push-forward transformations to the outcome space can be simulated using fitted model parameter estimates. 
Regardless of the specific modeling approach, experimenters should keep in mind that the value of the rational agent framework for gaining insight into a design or set of results depends on how well the behavioral scores predict expected performance in that experiment. Scores produced by a modeling approach that overfits to the particular observed behavior in the experiment (e.g., overfit to the particular combination of participants as shown in the example by~\cite{yarkoni2022generalizability}) will produce overfit benchmarks.

%% file: 02a-basic-framework.tex
\subsection{Decision Problems}
\label{sec:decision-problem}


Decision theory provides a natural framework for understanding
an agent's task in a visualization study.  A decision problem starts
by assuming a state space $\statespace$ that describes the set of
finite values (scenarios) that an uncertain state can take. Each
possible state $\state \in \statespace$ is a description of reality,
and only one may hold at a time. A \textit{data generating model}
defines a distribution over scenarios $\dist \in
\distover{\statespace}$.  In many experiments the distribution over
states is uniform.

A decision problem is defined by a distribution over states $\dist \in
\distover{\statespace}$ an action space $\actionspace$ and a {\em
  scoring rule} $\score : \actionspace \times \statespace \to \reals$
that maps the action and state to a quality or payoff.  Given a
distribution $\dist$ and scoring rule $\score$ denote the expected
score of an action by $\score(\action,\dist) =
\expect[\state\sim\dist]{\score(\action,\state)}$. The optimal decision
for a distribution $\dist$ is the one with the highest expected quality,
i.e.\ $\action^* = \argmax\nolimits_{\action \in \actionspace} \score(\action,\dist).$

In decision problems corresponding to prediction tasks, the action
space is a probabilistic belief over the state space, i.e.,
$\actionspace = \distover{\statespace}$.  For such problems, a
scoring rule is said to be \textit{proper} if the optimal action is to predict
the true distribution, i.e., $\dist = \argmax_{\action
  \in \actionspace} \score(\action,\dist)$.  Squared loss, a.k.a., the quadratic
scoring rule, is an example of a proper scoring rule that measures
the accuracy of beliefs.  For any scoring rule
$\score : \actionspace \times \statespace \to \reals$ there is an
equivalent \textit{proper scoring rule} $\proper : \distover{\statespace}
\times \statespace \to \reals$ defined by playing the optimal action 
under the reported belief.  Formally,
\begin{align}
  \label{eq:proper}
  \proper(\dist,\state) &= \score(\argmax\nolimits_{\action \in \actionspace} \score(\action,\dist),\state).
\end{align}


\paragraph{Example}


We illustrate the framework with a hypothetical weather forecast experiment, loosely inspired by \cite{savelli2013advantages}. Imagine a researcher who wants to compare people's performance in making a decision using several visualization strategies for presenting a predicted daily low temperature with uncertainty (i.e., a temperature distribution). They define a task in which the participant must decide whether to salt the parking lot or not,
i.e., by selecting action $\action$ from action space $\actionspace =
\{0=\text{no salt}; 1=\text{salt}\}$. They plan to score the participants for each decision task by simulating a temperature according to the predicted distribution.  The payoff relevant state
$\state$ is from state space $\statespace = \{0 = \text{not freezing},
1 = \text{freezing}\}$, corresponding to whether the simulated temperature was above or below the freezing point. Given the state space
$\statespace = \{0=\text{not freezing};1=\text{freezing}\}$ the
experimenter endows the following payoff function as a scoring rule:

\vspace{-2mm}
\begin{equation}
    \score(\action, \state)=\left\{\begin{array}{ccc}
    0 & \text{if }a = 0, \state= 0
      & \textit{no salt, not freezing } \\
     -100    & \text{if }\action=0, \state=1
     & \textit{no salt, freezing}\\
    -10      & \text{if }\action=1, \state= 0
    & \textit{salt,  not freezing}\\
    0    &  \text{if } \action=1, \state=1
    & \textit{salt, freezing}
    \end{array}\right.
    \label{eq:salting}
\end{equation}


\subsection{Information Structures and Visualizations}

\begin{table}[tbp]
\small
    \centering
    \begin{tabular}{c|c}
    \hline
     Payoff-relevant state    &   $\state\in \statespace$\\
      Signal (visualization) & $\signal \in \signalspace$\\
      Data generating process    & $\joint \in \distover{\signalspace
  \times \statespace}$\\
          Agent's action & $\action\in \actionspace$\\
      Scoring rule (payoff) & $\score : \actionspace \times \statespace \to \reals$\\
      \hline
     \end{tabular}
    \caption{Notation for defining a visualization experiment (assuming a single visualization strategy).}
    \normalsize
    \vspace{-6mm}
    \label{tab:rationa-agent-notation}
\end{table}


In a visualization experiment, the subject is given a stimulus in the
form of a visualization that is associated with the state.  Since the
visualization is associated with the state, if the subject understands the
visualization well, he can improve his performance at the decision
task.

To gauge the performance of a behavioral subject in such a task we
introduce the rational agent who faces the same task with the same
stimulus.  Formally, a visualization strategy induces an information
 structure that is given by a joint distribution $\joint \in
 \distover{\signalspace \times \statespace}$ over signals $\signal
 \in \signalspace$ (corresponding to the visualization) and states
 $\state \in \statespace$.  This joint distribution assigns to each
 realization $(\signal,\state) \in \signalspace \times \statespace$ a
 probability denoted $\joint(\signal,\state)$.
 The joint distribution allows us to calculate expected performance in the experiment. In the data generating process, there may be a fine-grained state $\finestate\in \finestatespace$ which determines the payoff-relevant state $\state$, i.e.\ there exists a function $\finestatemap$ that $\state = \finestatemap(\finestate)$. 

 Our framework allows us to study the performance of a single visualization strategy, or to compare a set of $k$ visualization strategies, inducing information structures $\joint_1, \joint_2, \dots, \joint_k$, respectively. 

\begin{figure}[htb]
\vspace{-2mm}
    \centering
    \includegraphics[width=3in]{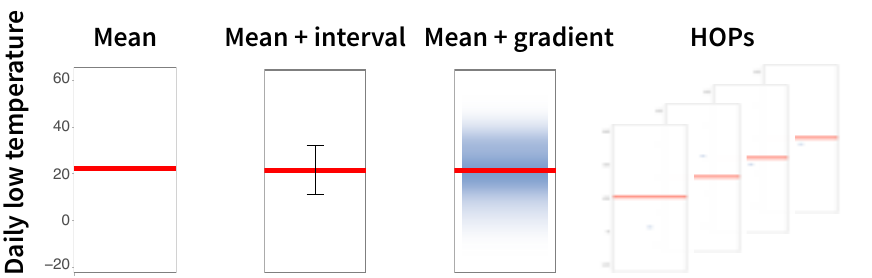}
    \caption{Example visualizations for a hypothetical weather forecast task.}
    \label{fig:weather_stimuli}
    \vspace{-2mm}
\end{figure}

 \paragraph{Example} 
The experimenter decides to evaluate a few different visualization strategies that can be used to present a weather forecast (Figure~\ref{fig:weather_stimuli}) for the decision problem they designed (\Cref{sec:decision-problem}). One shows only the expected daily low temperature. Another shows the expected low plus an interval expressing a 95\% confidence interval on the point estimate. Two others depict the probability distribution over possible low temperatures as a gradient plot (plotting probability as opacity) and animated hypothetical outcome plot (HOPs)~\cite{hullman2015hypothetical} (plotting probability as frequency). 


They define a data-generating process as follows: the daily low temperature
is generated from a Gaussian distribution $N(\mu, \sigma^2)$ with a
deterministic mean $\mu=5^\circ C$ and standard deviation
$\sigma$. The standard deviation $\sigma$ is uniformly drawn from
$\{2, 3, 4, 5\}$.  

For visualization strategies that depict uncertainty (CI,
gradient, HOPs), the signal $\signal$ is $(\mu,\sigma)$; for the visualization of the
mean, the signal $\signal$ is deterministically $\mu$.

The data-generating process results in a joint distribution $\joint \in \distover{\signalspace \times \statespace}$ on signal and state for the three non-trivial visualization strategies. The joint distribution allocates probability to getting a decision task for different combinations of $\theta$ and $\sigma$ in \Cref{tab:joint-weather-forecast}.

\begin{table}[htbp]
\vspace{-2mm}
    \centering
{    \small
\begin{tabular}{c|cccc}
\hline
      visualization $v$ for $\sigma$       & $\sigma = 2$ & $\sigma = 3$ & $\sigma = 4$ & $\sigma = 5$ \\
  \hline
  $\state = 0$  & $0.24845$ & $0.23805$ & $0.2236$ & $0.2103$ \\
  $\state = 1$  & $0.00155$ & $0.01195$ & $0.0264$ & $0.0397$ \\
  \hline
\end{tabular}
    \caption{The joint distribution $\joint\in \distover{\signalspace \times \statespace}$ on signal and state for the three non-trivial visualization strategies in the weather forecasting experiment. }
    }
    \label{tab:joint-weather-forecast}
    \vspace{-5mm}
\end{table}


The notation for the weather forecasting experiment is summarized in \Cref{tab:example-notation}.
\begin{table}[htbp]
\small
    \centering
    \begin{tabular}{c|l}
    \hline
     Payoff-relevant state    &   $\state\in \{0, 1\}$\\
     & \quad $=$ \{not freezing, freezing\}\\
     \hline
     Data generating model    
  & \tabitem fine-grained state: daily low temperature \\
  & \quad $\finestate\sim N(\mu, \sigma^2)$; $\state=\finestatemap(\finestate)=\indicator{\finestate\leq 0}$\\
  & \quad $\Pr[\state=1]=\Pr[t\leq 0]$;\\
  &  \quad $\mu=5$ fixed; \\
  & \quad $\sigma$ uniformly from $\{2, 3, 4, 5\}$.\\
  &\tabitem equivalently, \\
  &\quad $\Pr[\state=1]$ uniformly from\\
  &$0.62\%, 4.78\%, 10.56\%, 15.87\%$.\\
         \hline
         Agent's action & $\action\in \{0 = \text{no salt}, 1 = \text{salt}\}$\\
     \hline 
     Signal (visualization) & $\signal^{\text{vis}}\in V^{\text{vis}}$, vis $=$ visualization strategies\\
     &  vis $\in$ \{mean, CI, gradient,  HOPs\}\\
     & \qquad\quad of temperature\\
     \hline
     Scoring rule (payoff) & $\score(\action, \state)$ (see \cref{eq:salting})\\
     \hline
    \end{tabular}
    \caption{Notation for the freezing-salting example.}
    \normalsize
    \vspace{-5mm}
    \label{tab:example-notation}
\end{table}

The agent's belief about the freezing state $\state$ can be represented by the probability $\dist=\Pr[\state=1]$ of freezing. The corresponding proper scoring rule is 

\begin{equation}
    \proper(\dist, \state)=\left\{\begin{array}{ccc}
    0 & \text{if }\dist\leq 0.1, \state= 0
      & \textit{no salt, not freezing } \\
     -100    & \text{if }\dist\leq 0.1, \state=1
     & \textit{no salt, freezing}\\
    -10      & \text{if }\dist> 0.1, \state= 0
    & \textit{salt,  not freezing}\\
    0    &  \text{if } \dist> 0.1, \state=1
    & \textit{salt, freezing}
    \end{array}\right.
    \label{eq:salting-proper}
\end{equation}



\subsection{The Rational Agent: Baseline, Benchmark, and Information Value}


Two key constructs in our analysis of a behavioral agent are the
decisions of a rational agent without the visualization and with the visualization.  
In each case, the rational agent makes perfect use of the information available to them. In the case where they have access to a visualization, they do so by 
Bayesian updating from the joint distribution $\joint$ to a posterior
belief. Here we define the rational agent for a single visualization strategy. 

The rational agent's belief prior to the stimulus is their {\em prior
  distribution}:
\begin{align}
  \label{eq:prior}
\prior(\state) &= \sum\nolimits_{\signal \in \signalspace} \joint(\signal,\state). 
\intertext{The rational agent's belief after the stimulus is their {\em
    posterior distribution}.  The posterior belief is defined by following Bayes rule:}
\label{eq:posterior}
\posterior(\state) &= \joint(\state | \signal) = \tfrac{\joint(\signal, \state)}{\sum\nolimits_{\state \in \statespace} \joint(\signal, \state)}.
\end{align}

These two constructs induce a performance of the rational agent which
can be compared to the performance of the behavioral agent.  For a scoring rule $\score$ and information structure $\joint$, denote the corresponding proper scoring rule by $\proper$, prior distribution by
$\prior$, and posterior distribution by $\joint(\state|\signal)$.
Consider:

\begin{description}
\item[rational baseline:] The rational baseline is the performance of the rational agent without access to the signal, i.e., with only the prior belief.  
  \begin{align}
    \label{eq:rprior}
    \rprior &= \expect[\state \sim \prior]{\proper(\prior,\state)}.
\end{align}
\item[rational benchmark (visualization optimal)]
The rational benchmark is the performance of the rational agent with access to the signal, i.e., with the posterior belief.
\begin{align}
  \label{eq:rscore}
    \rscore &= \expect[(\signal,\state) \sim \joint]{\proper(\joint(\state|\signal),\state)}.
\end{align}
The expected payoff of any behavioral agent with the same visualization is below the rational benchmark. 

\item[value of information:] The difference between the rational benchmark  and the rational baseline quantifies the value of the information being visualized in the context of the scoring rule:
 \begin{align*}
     \infoval &= \rscore - \rprior.
 \end{align*}
The value of information provides a unit of difference in expected score for comparing behavioral performance. 
\end{description}

\subsubsection{Multiple Visualization Strategies}

When the framework is applied to multiple visualization strategies, the visualization optimal may vary. To compare multiple visualization strategies, the rational benchmark is defined with regards to the most helpful visualization.  Suppose the experimenter is comparing a set of $k$ different visualization strategies, with information structures $\joint^1, \dots, \joint^k$. 

\begin{description}
\item[visualization optimal:]
The visualization optimal is the performance of the rational agent with access to the signal, i.e., with the posterior belief.
\begin{align}
  \label{eq:rscore-general}
    \rscore^k &= \expect[(\signal,\state) \sim \joint^k]{\proper(\joint^k(\state|\signal),\state)}.
\end{align}
The expected payoff of any behavioral agent with the same visualization is below the visualization optimal. 
\item[rational benchmark:] Given multiple visualization strategies, the rational benchmark is instead defined as the best performance of the rational agent across different visualization strategies. 
 Suppose the experimenter aims to compare visualization formats $1\dots k$, inducing information structures $\joint^1, \dots, \joint^k$. The rational benchmark is defined as \begin{align}
  \label{eq:rperfect-general}
    \rperfect &= \max\nolimits_i\expect[(\signal,\state) \sim \joint^{i}]{\proper(\joint^{i}(\state|\signal),\state)}.
\end{align}
\end{description}

In addition to behavioral losses due to not properly receiving information or not optimizing one's decision (discussed below), we define an information loss induced by information asymmetry across visualizations, quantifying the extent to which visualization strategies provide varying amounts of information about the uncertain state.
\begin{description}
\item[information loss]
 The information loss captures the loss of information when data is summarized into a less informative visualization. We measure the information loss for a given visualization strategy by the difference $(\rperfect - \rscore) / \infoval$ between the rational agent benchmark (the rational best performance across visualizations) and the visualization optimal for a particular visualization strategy. 
\end{description}

\paragraph{Example}

We pre-experimentally analyze the hypothetical weather forecast experiment. 

We first calculate the prior and posterior distributions of the rational agent. Note that a distribution $\dist$ on a binary state space
$\statespace = \{0,1\}$ can be fully described by the probability that
the binary state is $\state = 1$ (freezing).  From
\cref{eq:prior} we have the prior probability of freezing $p=0.0796$.
and the posterior probabilities are $\Pr[\state=1|\sigma]=0.62\%, 4.78\%, 10.56\%, 15.87\%$, relatively for $\sigma=2, 3, 4, 5$, as given in \Cref{tab:example-notation}. 

\begin{figure}
    \centering
    \includegraphics[width=0.8\columnwidth]{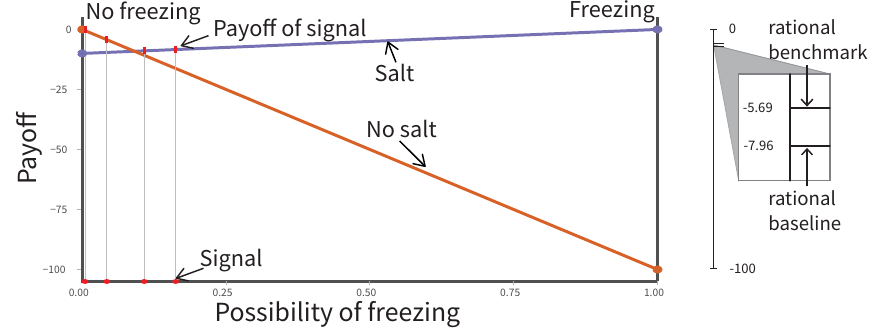}
    \vspace{-3mm}
    \caption{Score $\score(\action,\dist)$ as a function of belief $\dist \in [0,1]$ as probability of freezing.}
    \vspace{-5mm}
    \label{fig:weather_payoff}
\end{figure}

\Cref{fig:weather_payoff} depicts the expected score of the agent for both no-salt and salt actions as a function of her belief
$\dist$, as specified in \Cref{eq:salting}.  Notice that if the belief is certainty either $0$ or $1$,
then the payoff is given explicitly by the scoring rule.  For an
uncertain belief $\dist \in (0,1)$ between $0$ and $1$ the payoff is given by linearly
interpolating between certain beliefs, i.e., the payoff is the
expected value of the action over the belief.  Lines correspond to the no-salt and salt action.  The optimal
action for each posterior belief -- i.e., the action taken by the
rational agent -- can be read off as well.  For each
signal, we find its posterior on the horizontal axis, and evaluate
which of the two actions give a higher payoff and take that one.  From
this analysis it is clear that the no-salt action $\action = 0$ is
taken on the lower two signals $\{2,3\}$ and the salt action
$\action=1$ is taken on the higher two signals $\{4,5\}$. The payoff
lines cross at $\dist = 0.1$ where the decision-maker is indifferent
between no-salt and salt actions, so the proper scoring rule in \Cref{eq:salting-proper} sets belief threshold at $\dist=0.1$.

The rational agent framework gives the following quantities:
\begin{description}
\item[rational baseline:] $\rprior = -7.96$.  

  The prior $\prior = 0.08$ is optimized at no-salt and gives an expected payoff of $-7.96$. 

The calculation is as following:
\begin{align*}
    \rprior &= \Pr[\state=0]\cdot\score(\action = 0, \state=0) + \Pr[\state=1]\cdot\score(\action = 0, \state=1)\\
    &= (1-0.0796)\times 0 + 0.0796\times (-100)=-7.96
\end{align*}

\item[visualization optimal:] 
  $\rpos^{\text{CI}}=\rpos^{\text{gradient}}=\rpos^{\text{HOPs}}=-5.69$; $\rpos^{\text{mean}}=-7.96$.
  
  In CI, gradient, and HOPs, each signal arises with probability $1/4$ and the average of the optimal actions under the induced posteriors (read off \Cref{fig:weather_payoff}) gives $\rpos = -5.69$.  For the visualization of the mean, the rational agent has only the prior information and obtains $\rpos^{\text{mean}} = \rprior = -7.96$.

  The calculation of $\rpos$ for CI, gradient, and HOPs is the following:
  \begin{align*}
      \rpos^{\text{CI}} &= \sum\nolimits_{\sigma, \state}\Pr[\sigma, \state]\cdot \score(\mathbf{1}_{\sigma\in\{4, 5\}}, \state)\\
      &= 0.24845\times 0 + 0.00155\times (-100)+0.23805\times 0 + 0.01195\times (-100)\\
      &\quad + 0.2236\times (-10) + 0.0264\times 0 + 0.2103\times (-10) + 0.0397\times 0\\
      &= -5.69
  \end{align*}

 \item[rational benchmark:] $\rperfect = \max_{\text{vis}} \rscore^{\text{vis}} = -5.69$, the best achievable across visualizations. 
\item[value of information:]  $\infoval = \rperfect - \rprior = 2.27$.

  \end{description}
     

 Suppose the experimenter sets the conversion rule $f(\reward) = \$1+\$0.01\reward$ from score $\reward$ to real dollars as follows: an agent gains a fixed $\$1$ for completing each trial,  plus a $\$0.01$ in real dollars for each point earned in scoring rule space.  The conversion rule is set such that an agent is guaranteed to obtain a positive payment. We calculate the expected real payments to a rational agent in  \Cref{tab:forecast-incentive}. If the goal is to incentivize an agent to consult the visualization, we would conclude that the incentive is badly designed because it is a very small fraction of the amount expected without looking at the visualizations (<3\%).

\begin{table}[htbp]
\vspace{-2mm}
\small
    \centering
    \begin{tabular}{c|c|c|c}
    \hline
     $f(\rprior)$ & $f(\rpos)$ &  $\Delta_f$ & $\Delta_f / f(\rprior)$\\
     \hline
     $\$0.920$ & $\$0.943$ & $\$0.023$ & $2.5\%$\\
    \hline
    \end{tabular}
    \caption{$f(\rprior)$ shows the expected payment to a rational agent without the visualization, $f(\rpos)$ shows the expected payment to a rational agent who reads the visualization, while $\Delta_f=f(\rpos)-f(\rprior)$ is the incentive to consult the visualization.}
    \normalsize
    \vspace{-2mm}
    \label{tab:forecast-incentive}
\end{table}

The information loss can also be calculated pre-experimentally.
\begin{description}
    \item[information loss]  CI, gradient, and HOPs: $(\rperfect-\rpos)/\infoval = 0$. 
    
    Mean: $(\rperfect - \rpos^{\text{mean}})/\infoval = 100\%$. 
\end{description}

From this pre-experiment analysis, the experimenter should also expect the mean visualization to behave badly in payoff compared to the interval, because the mean has a information loss of $100\%$, i.e.\ it is not informative for the decision task. 



\subsection{The Behavioral Agent and Performance Analysis}

The behavioral agent faces the same task as the rational agent upon
seeing a visualization and choosing an action $\action$ from an action
space $\actionspace$. Once the experiment has been conducted the
collected data implies an empirical joint distribution
$\bjoint\in\distover{\actionspace \times \statespace}$ over the
behavioral actions and the states.

Experimenters can estimate the following measures to quantify behavioral performance: 
\begin{description}
    \item[behavioral score:] The behavioral score is the expected score of the behavioral agent.
      \begin{align}
        \label{eq:bscore}
\bscore &= \expect[(\action,\state) \sim \bjoint]{\score(\action,\state)}.
\end{align}
   \item[behavioral value of information:] The behavioral value of information is the difference between the behavioral score and the rational baseline (if non-negative).
    \begin{align*}
   \binfoval &= \max(\bscore - \rprior,0).
   \end{align*}
\end{description}

The behavioral score $\bscore$ is always below the rational benchmark
$\rscore$ and can be either above or below the rational baseline
$\rprior$.  Importantly, if the behavioral score is below the rational
baseline, then from the scores alone we cannot reject the hypothesis
that the behavioral agent got no useful information from the
visualization.  Even with no information, the rational agent performs
better.  On the other hand, if the behavioral score exceeds the
rational baseline, then the behavioral agent systematically performs
better than the rational agent with no information and, therefore,
must be getting some useful information from the visualization.

To understand how much useful information the behavioral agent is able
to get from the visualization, we consider the ratio of the value of information to
the behavioral value of information, i.e., $\binfoval / \infoval \in
[0,1]$.  If this ratio is large, i.e., close to one, then there is
little room to improve the amount of effective communication of the
visualization for the decision problem. If this ratio is small, then
there is theoretically an opportunity to improve communication. 

\subsection{Calibrated Behavior and Fine-grained Analysis}

The source of behavioral errors can be identified by observing that
the joint distribution of behavior and state may contain information
that the agent was not able to appropriately act on.  In other words,
the correlation between behavior and state captures information that is not necessarily reflected by the payoff. The agent's behavior may not be calibrated. The agent's behavior is calibrated
if action $\action \in \actionspace$ is the optimal action on the
conditional distribution over states when that action $\action$ was
taken.  The following calibrated behavioral score is always between
the rational baseline and the rational benchmark:

\begin{description}
    \item[calibrated behavioral score] The calibrated behavioral score is the score of a rational agent on information structure $\bjoint$.
      \begin{align}
        \label{eq:cscore}
    \cscore &= \expect[(\action,\state) \sim \bjoint]{\proper(\bjoint(\state | \action),\state)}.
\end{align}
\end{description}

The calibrated behavioral agent performance allows for different
behavioral errors to be distinguished, and the information conveyed by
the visualization to be assessed even when the behavioral score is
below the rational baseline.  We identify two sources of
loss for the behavioral agent:

\begin{description}

\item[belief loss] The belief loss captures the loss in score as a result of the agent not responding with different beliefs after looking at visualizations of informationally distinct stimuli (e.g., different proportions, probabilities, etc.). We measure the belief loss by calibrating the behavioral decisions and responses. The difference $(\rscore-\cscore) / \infoval$ quantifies the magnitude to which the agent is not able to differentiate between stimuli. 


\item[optimization loss] Upon viewing a visualization the rational agent would update their beliefs and then choose the optimal action under those beliefs. The optimization loss 
captures the loss from the agent not properly updating their beliefs about the uncertain state and making the optimal decision given their beliefs.  The difference $(\cscore - \bscore) / \infoval$ quantifies the magnitude to which the agent is unable to use the information they have obtained.
\end{description}

%% file: 03_demonstrations.tex
\section{Demonstrations}
We apply the rational agent framework to two visualization experiments. 
\footnote{See "demonstrations/effect\_size/analysis.Rmd" and "demonstrations/transit\_decisions/analysis.Rmd" in our supplementary material for the complete analysis. Our supplement is available at \url{https://github.com/Guoziyang27/rational\_framework}}
Both experiments won awards for their rigorous design at the conferences at which they were published, making them a conservative choice for demonstrating the interpretive value added by the framework. 

\input{03_1_kale}

\input{03_2_fernades}

%% file: 03_1_kale.tex
\subsection{Effect size judgments and decisions~\cite{kale2020visual}}
Kale et al.~\cite{kale2020visual} use an online crowdsourced experiment to investigate the extent to which visualization design impacts people's use of heuristics based on the central tendency in judging effect size~\cite{coe2002s}, a measure of the ``signal'' in a distributional comparison relative to the noise.

\subsubsection{Experiment design}
\begin{table}
\small
    \centering
    \begin{tabular}{c|l}
    \hline
     Payoff-relevant state    & \tabitem $\state_0\in \{0, 1\}$ \\
     & \quad = lose/win w/o.\ a new player\\
     & \tabitem $\state_1\in \{0, 1\}$\\
     & \quad = lose/win w.\ a new player\\
     \hline
     Data generating model   
  & \tabitem  fine-grained state $(\finestate_0, \finestate_1)$, where\\
  &\qquad $\finestate_0\sim N(100, \sigma^2)$\\
  & \quad = score w/o.\ a new player\\
   & \qquad $\finestate_1\sim N(\mu, \sigma^2)$\\
   & \quad = score w.\ a new player\\
   & \tabitem win: score higher than $100$, $\state_i=\indicator{\finestate_i\geq 100}$\\ 
   & \quad $\Pr[\state_i=1]=\Pr[\finestate_i\geq 100]$\\
  & \tabitem $\Pr[\state_0=1]=50\%$\\
  & \tabitem $\Pr[\state_1=1]$ uniformly drawn \\
  & from $\{p_1, \dots, p_8\}$\\
     \hline 
     Signal (visualization)& $v\in V$ visualizing  $\finestate_0, \finestate_1$ \\
     &  \quad e.g.\ CI, HOPs, densities, QDPs\\
         \hline
         Agent's action & $\action\in \{0 = \text{not hiring}, 1 = \text{hiring}\}$\\
     \hline
     Scoring rule (payoff) & $\score(a, \state)$\\
     \hline
    \end{tabular}
    \vspace{-2mm}
    \caption{Kale et al.\cite{kale2020visual} decision problem under our framework.}
    \normalsize
    \label{tab:kale-notation}
    \vspace{-6mm}
\end{table}

Kale et al.'s mixed design experiment compares judgments and decisions across four approaches to visualizing a pair of distributions: quantile dotplots (QDPs)~\cite{kay2016ish}, hypothetical outcome plots~\cite{hullman2015hypothetical}, 95\% containment intervals, and density plots, 
assigned between subjects. Each participant does trials where the means are visually annotated and where they are not. 
The distributions are framed as predicted scores in a fantasy sports game for a team with and without a new player. Participants are tasked with using the visualizations for a binary decision task: whether to pay to add the new player to their team, knowing that doing so increases their chance of winning a monetary award but costs money. Additionally, on each trial an unincentivized probability of superiority (PoS) judgment is elicited, representing the participant's belief about the probability that a random draw from the score distribution with the new player will be greater than one from the distribution without. 
This allows us to calculate belief and optimization loss for both a belief and a decision question.


 
\paragraph{Scoring rule}
\Cref{tab:kale-notation} summarizes the decision problem under our framework. The action space is $A=\{0, 1\}$ for the participant or equivalently $A=\{\text{not hire}, \text{hire}\}$. There are two fine-grained random states, one $\finestate_0$ indicating the score without a new player, and the other one $\finestate_1$ indicating the score with a new player. The agent wins a game if the realized score is above $100$, i.e. $\theta_i=\indicator{\finestate_i\geq 100}$.  The payoff function is defined by 
\begin{equation*}
    \score (a, \state)=\left\{\begin{array}{ccc}
    0 & \text{if }a = 0, \state_0=0
      & \textit{lose without hiring} \\
     3.17    & \text{if }a=0, \state_0=1
     & \textit{win without hiring}\\
    -1      & \text{if }a=1, \state_1=0
    & \textit{lose with new player}\\
    2.17     &  \text{if } a=1, \state_1=1
    & \textit{win with new player}
    \end{array}\right.
\end{equation*}
where the unit is millions of dollars in the simulated account. The simulated accounts are initialized with $108$M dollars. At the end of the experiment, the agents are rewarded $\$0.8$ per $1$M more than $150$M in their simulated accounts. 


\paragraph{Stimuli generation and optimal decision strategy}
The probability $\Pr[\state_0=1]$ of winning without a new player is fixed at $50\%$. The experiment varies the probability $\Pr[\state_1=1]$ of winning with a new player at $8$ levels above $50\%$, corresponding to 8 ground truth PoS sampled in log space from 0.55 to 0.95. 
The score  $\finestate_0$ and $\finestate_1$ follow a Gaussian distribution with identical standard deviations of either 5 or 15. $\finestate_0$ has a mean fixed at 100; the target PoS for each trial is realized by varying the mean of $\finestate_1$. 
Each block of trials the participant completes presents these eight levels twice, once with the lower standard deviation and once with the higher standard deviation.

The realized score in the fictional sports game (used to determine the participant's payoff for a trial) is simulated using Monte Carlo method. The agent faces a  decision problem of hiring the new player or not, where his expected utility is as follows:
\begin{align*}
    3.17\cdot\Pr[\finestate_0\geq 100] &\qquad\text{ if he does not hire;}\\
    2.17 \cdot \Pr[\finestate_1\geq 100]+(-1)\cdot\Pr[\finestate_1<100]&\qquad\text{ if he hires}.
\end{align*}
When the rational agent believes that $3.17\cdot\Pr[\finestate_1\geq 100]\leq 2.17 \cdot \Pr[\finestate_1\geq 100]+(-1)\cdot\Pr[\finestate_1<100]$, or equivalently that $\Pr[\finestate_1\geq 100]\geq 81.5\%$, her optimal decision is to choose to hire a new player and vice versa. 

As mentioned above, on each trial behavioral agents are asked for an unincentivized PoS judgment $\Pr[\finestate_1\geq \finestate_0]$. Under the choice to fix the mean of $\finestate_0$ at $100$, the PoS judgment maps to a unique probability of winning with a new player, thus mapping to a unique optimal decision.
As a result, the PoS judgment represents beliefs associated with the incentivized decision. 


\paragraph{Rational Agent} On any given trial, the agent is presented with a probability $\Pr[\finestate_1\geq 100]$ of winning with a new player, randomly drawn from the $8$ predetermined levels  $p_1, p_2, \dots, p_8$. Without getting any additional information (i.e., seeing any visualizations), the rational agent has prior belief $\Pr[\finestate_1\geq 100]=\frac{1}{8}\sum_{i=1}^8 p_i = 80.5\%$, so the optimal decision is always not to hire a priori. 

The rational agent knows the distributions of scores shown in the visualization follow Gaussian distributions which are parameterized by mean and variance. Different visualization strategies have the same value to the rational agent, regardless of whether means are added or not\footnote{A rational agent will spend infinite time looking at HOPs, to fully understand the distribution of scores.}. 
Hence, any visualization in the experiment is equivalent for the rational agent to show the probability of the team winning with the new player. After seeing the visualization, the rational agent knows $\Pr[\finestate_1\geq 100]=p_i$ for some $i$, and makes the optimal decision. 

\begin{figure}[t]
\centering
    \includegraphics[width = 0.6\columnwidth]{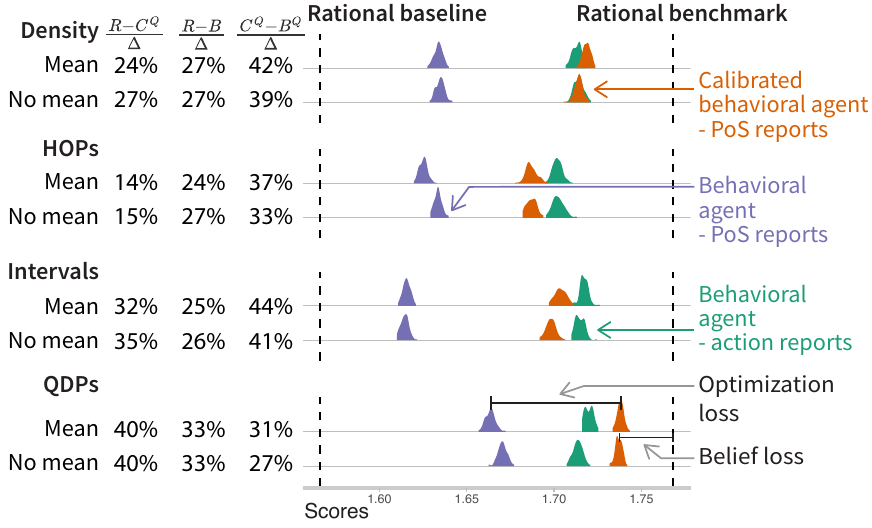}
    \caption{Estimated payoffs under the scoring rule used in Kale et al.~\cite{kale2020visual} for 100 simulated experiments in which behavioral agents make decisions (\textbf{\behavioralactiontextcolor{behavioral decision score}} $\bactionraw$, green) and report PoS judgments (\textbf{\behavioraltextcolor{PoS raw score}}, purple, and adjusted \textbf{\calibratedtextcolor{calibrated PoS score}}, orange) by visualization condition with means added and without. The rational agent benchmark $\rpos$ and the rational agent baseline $\rprior$ are shown as dotted lines. 
    }
    \vspace{-5mm}
    \label{fig:kale2021}
\end{figure}

Dotted lines in \Cref{fig:kale2021} show the rational baseline ($\rprior$, left) and rational benchmark ($\rpos$, right).

\subsubsection{Pre-experimental Analysis}

We calculate the rational agent baseline and benchmark for a single decision task, in simulated account dollars in millions.
\begin{description}
    \item[Rational baseline:] $\rprior = 1.57$. The rational agent achieves $\rprior$ by selecting any fixed action, or arbitrarily randomizing over the actions. 
    \item [Rational benchmark / visualization optimal:] $\rpos = 1.77$ for all visualization formats. 
    \item[Value of information:] $\Delta = \rpos - \rprior = 0.20$.
\end{description}

The information loss is $0$ for all visualization strategies.

When we translate these scores through the conversion rate to real dollars received by the participant ($f(r)=\$1 + \max\{0, \$0.08(r-150\text{M})\}$ for each $1$M over $150$M in the  account where $r$ is in millions), we get the total incentive that an agent has to consult the visualization, 
shown in \Cref{tab:kale-incentive}\footnote{With high probability, the simulated payoff falls over $150$M. $f$ can be considered linear here, so we write the expected real payment as $f(\rpos)$.}. This incentive seems reasonable for encouraging agents to consult the visualization, as it is nearly a third of the guaranteed payment from choosing any fixed action.

\begin{table}[htbp]
\small
    \centering
    \begin{tabular}{c|c|c|c}
    \hline
     $f(\rprior)$ & $f(\rpos)$ &  $\Delta_f$ & $\Delta_f / f(\rprior)$\\
     \hline
    $\$1.66$ &  $\$2.17$ & $\$0.51$ & $30.72\%$\\
    \hline
    \end{tabular}
    \caption{$f(\rprior)$ shows the expected payment to a rational agent, $f(\rpos)$ shows the expected payment to a rational agent who reads the visualization, while $\Delta_f=f(\rpos)-f(\rprior)$ is the incentive to consult the visualization.}
    \normalsize
    \vspace{-2mm}
    \label{tab:kale-incentive}
\end{table}


One point worth acknowledging is that Kale et al.\ do not provide participants with the prior, as is frequently true in visualization experiments. This is not necessarily a flaw in the design. In this example, there are reasons why we would expect behavioral agents to achieve scores higher than $\rprior$ in the experiment design despite not explicitly being given the prior. 
For this example, the prior score can be obtained by taking the same action in any trial or arbitrarily randomizing over actions. 
Additionally, participants were given feedback, and a participant who was randomizing but watching feedback is arguably in a position to approximately learn the prior over the course of the experiment.




\subsubsection{Post-experimental Analysis}

The original results presented by Kale et al.~\cite{kale2020visual} include a consistent but very small impact of annotating means on bias in PoS judgments, and some disparity between what visualizations appear to perform best for PoS judgments versus incentivized decisions: QDPs perform relatively well across the two tasks, but performance with intervals and densities varies across tasks. 
The authors advise visualization researchers to be cautious in assuming that perceptual accuracy feeds directly into decision-making,
because a user’s internal sense of effect size is not necessarily identical when they use
the same information for different tasks.
The authors speculate that the decoupling of performance may result from users relying on different heuristics
to judge the same data for different purposes. (e.g., Kahneman and Tversky’s~\cite{kahneman2013prospect} suggestion of a distinction between perceiving an event's probability and weighting the probability in decision-making), or from not incentivizing the PoS question.
By applying the rational agent framework post-experimentally, we further investigate their results and this ambiguity.

In our post-experimental analysis, we first empirically estimate the expected payoff $\bactionraw$ for decisions. 
Because the study hypothesis in Kale et al.\ concerned the comparison between performance with means annotated versus not annotated, we calculate 
the expected \textbf{\behavioralactiontextcolor{behavioral score for the decision task}} for each of the four visualization strategies crossed with the means manipulation, resulting in eight total scores with uncertainty (\Cref{fig:kale2021}, \behavioralactiontextcolor{\textbf{green}}).

Specifically, we calculate these scores by simulating
binary decisions for the intended number of agents per combination of visualization approach and means manipulation (of eight) in the original experiment (160 people per visualization approach, each of which completed a block of 16 task trials with and without means).\footnote{In reality, less than 160 were achieved for some conditions in the original experiment. Replicating the missing data structure instead of using the intended cell count does not change our results.} 
For each condition we repeatedly sampled $n=160\times 16$ simulated responses from the posterior predictive distribution of the Bayesian logistic regression model used by Kale et al.~\cite{kale2020visual}, balancing trial numbers and block orders according to the original experiment design. We report scores obtained from simulating results $100$ times (\Cref{fig:kale2021}, \behavioralactiontextcolor{\textbf{green}}).
These scores indicate that the behavioral agents' decisions achieved a payoff higher than the rational agent with prior and fairly close to the rational agent with posterior, which we further analyze below. 


Kale et al.\ \cite{kale2020visual} elicit responses on a finer space $\responsespace=\Delta(\{0, 1\})$ - the PoS reports, which is more informative than their decision task in that each PoS corresponds to a unique belief on the winning probability.  We  apply our framework  by calculating the scores from PoS reports. To calculate expected \textbf{\behavioraltextcolor{behavioral scores $\bresponseraw$ 
for the PoS task}}, we simulate decisions by applying the optimal decision rule to reported PoS, however this time we sampled from the posterior predictive distribution of the authors' linear-in-log-odds model for PoS judgments (\Cref{fig:kale2021}, \textbf{\behavioraltextcolor{purple}}). 
Scores for the PoS task are closer to the prior than those for the decision task.
Similar to Kale et al.'s results, for both the decision task and PoS task we see only a slight difference in expected behavioral scores with and without the addition of means.

Finally, we calculate the calibrated behavioral scores.  The calibrated scores for decisions are the same as the expected payoff $\bactionraw$; recall this is because for a binary decision where the behavioral score is above $\rprior$, calibration cannot improve the score. 
We follow the same  approach to calibrate PoS reports and calculate the \textbf{\calibratedtextcolor{calibrated behavioral scores $\bresponsecalib$ for the PoS task}} by 
discretizing the PoS report space (\Cref{fig:kale2021}, \calibratedtextcolor{\textbf{orange}}). We discretize the space into intervals of length $0.02$ so that we can calculate the empirical Bayesian posterior of state $\state_1$ without overfitting.\footnote{Note that discretization induces an unavoidable discretization error to the estimation of calibrated score.} 

\paragraph{Belief Loss}
Recall that belief loss measures the extent to which a behavioral agent can distinguish between stimuli by consulting the visualization, and is quantified by taking the difference between the rational benchmark and the calibrated behavioral responses, $\rpos - \cscore$, and normalizing by $\Delta$.  Because calibrating the decision scores does not improve upon the behavioral scores for Kale et al's decision task, belief loss is equivalent to $\frac{(\rpos - \bscore)}{\Delta}$ in \Cref{fig:kale2021}. 



We next consider belief loss for the PoS task as $\frac{\rpos-\bresponsecalib}{\Delta}$ in \Cref{fig:kale2021}. 
QDPs induce the least belief loss and HOPs the most. This may be because agents will often not watch the HOPs animation for long, and hence are lossy information processors compared to the rational agent~\cite{kale2020visual}.  
The ranking we observe across visualization conditions resembles that observed in the Just-Noticeable-Difference (JND) estimates in Kale et al.'s model of participants' decisions. JNDs measure how sensitive behavioral agents are to the evidence in making decisions. 

\paragraph{Optimization Loss}
Recall that optimization loss is calculated as $\frac{(\cscore - \bscore)}{\Delta}$. This loss is 0 for the decision task because expected scores were above $\rprior$. 
When we evaluate optimization loss for the PoS task, we observe fairly substantial gaps between the behavioral and calibrated behavioral scores (\behavioraltextcolor{\textbf{purple}} and \calibratedtextcolor{\textbf{orange}} distributions). The normalized optimization loss is shown as $\frac{\bresponsecalib-\bresponseraw}{\Delta}$ in \Cref{fig:kale2021}. 
These scores indicate 1) that the behavioral agents are struggling to report their beliefs but getting information from the visualizations, and 2) the 
 behavioral agents are getting a fair amount of information from the visualizations: the calibrated scores are obtaining a relatively high percentage of the rational benchmark. 

 When we look at decision scores, and compare them to calibrated PoS, we see that the behavioral agents are making nearly optimal decisions given the information they have (to hire the new player or not). This is because we can expect the PoS reports to capture the agents' perceived probability of winning with the new player (due to the one-to-one mapping between PoS and probability of win by design). 
 This suggests agents are understanding the experiment task fairly well.


 The fact that behavioral scores for the PoS report are considerably improved by calibrating indicates that agents struggled to use the information they had obtained to report their beliefs.
 Kale et al. acknowledge that they cannot disambiguate the reason for the disparity in the PoS versus decision results they observe, and speculate it may stem from the PoS question not being incentivized or from a difference between probability perception and weighting~\cite{kahneman2013prospect}. However, our comparison between expected scores for the binary decision task and the PoS task suggests that agents \textit{were} consulting the visualizations and extracting much of the information.

Alternative reasons agents may have struggled with reporting for the PoS question is that while Kale et al.'s design cleanly maps PoS to probability of winning with the new player, the latter is the more directly relevant information to the decision at hand. 
PoS is also harder to read from the visualizations that the participants were provided relative to the probability of winning.
Our analysis calls into question the possible explanations proffered in the paper for explaining differences observed in how visualizations perform between PoS and decision tasks.
Had the experiment asked a directly payoff-related question like \textit{What is the improvement in the probability of winning by hiring a new player?} the comparison the work makes between beliefs and decisions may have been more informative for assessing conjectures like Kahneman and Tversky's notion of differences in probability perception and weighting~\cite{kahneman2013prospect}.

%% file: 03_2_fernades.tex
\subsection{Transit decisions~\cite{fernandes2018uncertainty}}

\begin{figure}
   \centering
  \includegraphics[width=0.6\columnwidth]{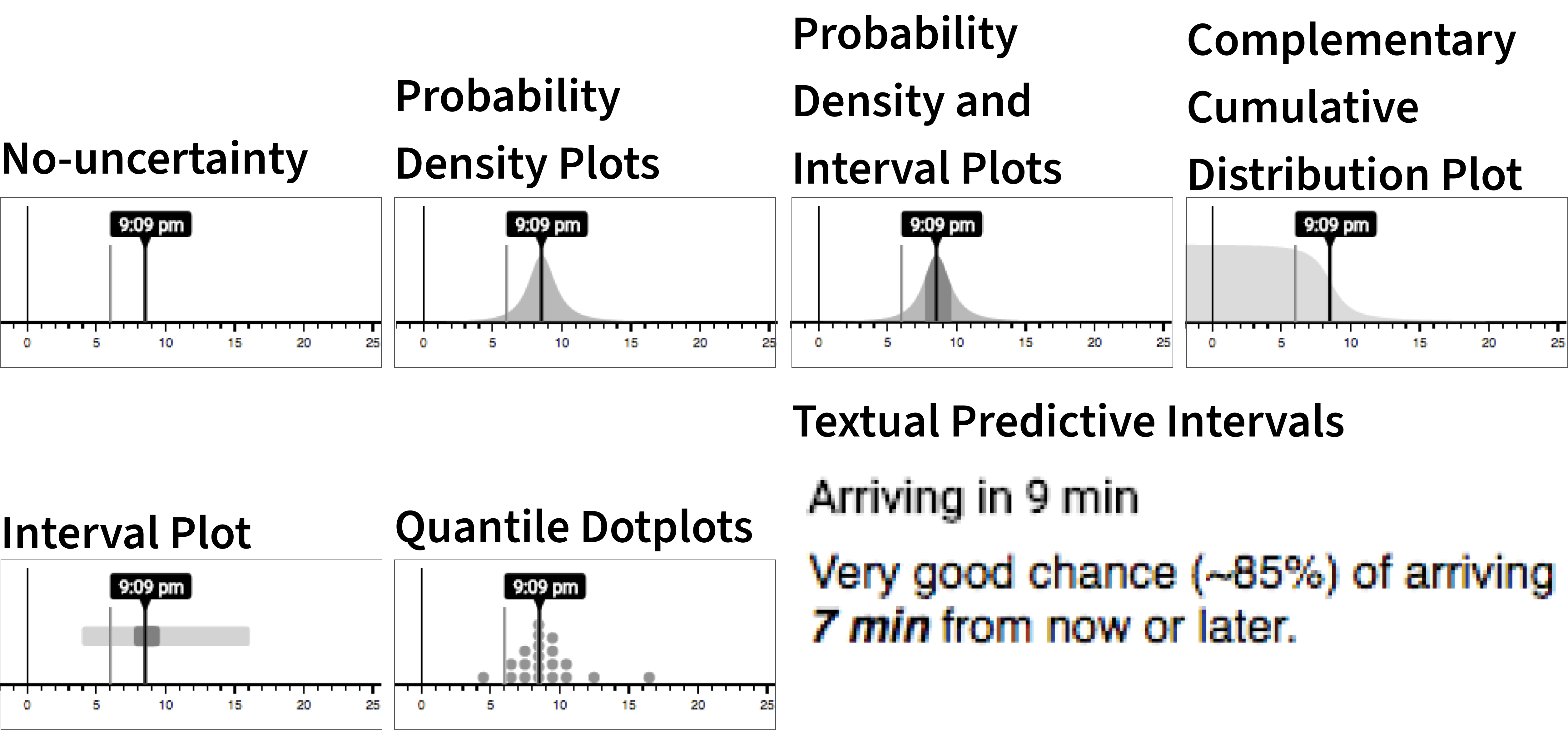}
 \caption{Stimuli from Fernandes et al. \cite{fernandes2018uncertainty}}
    \label{fig:fernandes_stimuli}
\end{figure}

Fernandes et al.~\cite{fernandes2018uncertainty} compare different approaches to presenting bus arrival time predictions--including textual descriptions of one-sided probability intervals, containment intervals, QDPs, CDFs, density plots, density plots with intervals, and only a point estimate (no uncertainty control)--for making transit decisions about when to leave for the bus stop.  

\subsubsection{Experiment Design}

\begin{table}[htbp]
\vspace{-3mm}
    \small
    \centering
    \begin{tabular}{c|l}
    \hline
     Payoff-relevant state    & $\state\in [0, 30]$ bus arrival time\\
     \hline
     Data generating model
   & $\state$ from Box-Cox $t$ distribution\\
     \hline 
     Signal (visualization)& $\signal\in \signalspace$ visualizing  $\state$ \\
         \hline
         Agent's action & $\action\in [0, 30]$ time to go to bus stop \\
     \hline
     Scoring rule (payoff) & $\score(\action, \state)$\\
     \hline
    \end{tabular}
    \caption{Decision problem for Fernandes et al.\cite{fernandes2018uncertainty}}
    \normalsize
    \vspace{-3mm}
    \label{tab:fernandes-notation}
\end{table}

Fernandes et al.'s mixed design experiment compares incentivized decisions across twelve visualization strategies that are assigned between subjects. Each participant is presented with 40 total trials parameterized by bus arrival time distributions.
Participants are randomly assigned one of three decision scenarios representing a hypothetical real-world decision with an associated (unique) scoring rule. 

The decision problem is summarized in \Cref{tab:fernandes-notation}. The agent takes action from $\actionspace=[0, 30]$, a time to arrive at the bus stop. The payoff-relevant state is $\state\in[0, 30]$, the time the bus arrives at the bus stop. When $\action>\state$, the agent does not catch the bus. If he misses the bus, he is guaranteed to catch a second bus that arrives at $\state'+30$, where $\state'$ follows the same arrival distribution as the first bus. In each of the three decision scenarios, the agent gains a bonus $\reward_0>0$ for each minute of activities before arriving at the bus stop, $\reward_w<0$ for each minute waiting at the bus stop, and a bonus $\reward_d>0$ for each minute spent at the destination with a maximum time of $\maxtime$ spent.  The payoff can be formulated as follows:
\begin{equation}
    \score(\action, \state)=\left\{\begin{array}{cc}
    \reward_0\action + \reward_w(\state-a) + \reward_d \cdot \maxtime     & \text{if }\action\leq \state\\
   \text{catching bus}\\
    \reward_0\action + \reward_w(\state'+30 - \action)+\reward_d\cdot [\maxtime-(\state'-\state)]     & \text{else}\\
    \text{not catching bus}
    \end{array}\right.
\end{equation}
For each decision scenario, payoffs are generated as in \Cref{tab:fernades-payoff}.
\begin{table}[htbp]
\small
    \centering
    \begin{tabular}{c|c|c|c|c}
    \hline
    Scenario ID     &  $\reward_0$ & $\reward_w$ & $\reward_d$ & \maxtime\\
    \hline
     1    & 8 & -14 & 14 & 90\\
         \hline
         2 & 14 & -14 & 14 & 60\\
         \hline
         3 & 8 & -17 & 17 & 120\\
         \hline
    \end{tabular}
    \vspace{-2mm}
    \caption{Payoffs of decision tasks for different scenarios. }
    \normalsize
    \vspace{-3mm}
    \label{tab:fernades-payoff}
\end{table}

\paragraph{Stimuli generation and optimal decision strategy} 
Each trial corresponds to a Box-Cox $t$ distribution generated from a model of real bus arrival predictions~\cite{kay2016ish}. Fixing a belief distribution $\dist$ where the arrival time $\state$ is drawn, if the agent chooses action $\action$, his expected payoff is
\begin{align}    
    \E_{\state\sim \dist}[\score(\action, \state)]=\sum_{\state\leq \action}\Pr[\state]\left[ \reward_0\action  + \reward_w(\state-\action) + \reward_d \cdot \maxtime \right]\nonumber\\
    + \sum_{\state>\action} \Pr[\state]\left[\reward_0\action + \reward_w(\E_{\state'\sim \dist}[\state']+30 - \action)+\reward_d\cdot [\maxtime-(\E_{\state'\sim \dist}[\state']-\state)]\right].
\end{align}


\paragraph{Rational Agent} 

The visualizations are informationally equivalent to the rational agent and equivalent to knowing the bus arrival distribution, except for the text displays. 
This is because, with the exception of text displays, there is a one-to-one mapping between the distribution visualization on a trial and the bus arrival distribution. Note that this is also true for no uncertainty displays (control).
The no uncertainty condition visualization displays the mean of the bus arrival distribution. Each bus arrival distribution in the experiment has a distinct mean, so the rational agent fully knows the bus arrival distribution after seeing the mean. After seeing the visualization, the rational agent knows the bus arrival distribution $D$, thus is able to make the optimal decision. 
For the text probability interval displays, however, 
the rational agent is not able to distinguish between distributions that map to the same text, leading to a lower expected score. 


\subsubsection{Pre-experimental Analysis}\label{sec:fernandes-preexperimental}


We calculate the rational agent baseline, visualization optimal, and rational benchmark for a single trial in the unit of simulated coins. 



\begin{description}
    \item[Rational baseline:] 
    \begin{table}[htbp]
    \vspace{-2mm}
\small
    \centering
    \begin{tabular}{c|c|c|c}
    \hline
     Scenario ID & 1 & 2 & 3\\
    \hline
    $\rprior$ & 1078.7 & 767.5 & 1850.2\\
    \hline
    \end{tabular}
    \vspace{-2mm}
    \caption{The rational baseline $\rprior$ for different scenarios.}
    \normalsize
    \vspace{-4mm}
    \label{tab:fernandes-prior-score}
\end{table}
\Cref{tab:fernandes-prior-score} summarizes the baseline $\rprior$. The rational agent achieves $\rprior$ by selecting a fixed action. 
    \item[Visualization optimal:] 
    \Cref{tab:fernandes-posterior-score} summarizes the visualization optimal $\rpos$.
    \begin{table}[htbp]
    \vspace{-2mm}
\small
    \centering
    \begin{tabular}{c|c|c|c}
    \hline
     Scenario ID & 1 & 2 & 3\\
    \hline
    $\rpos$ full information & \multirow{2}{2.8em}{1171.8} & \multirow{2}{2.8em}{852.0} & \multirow{2}{2.8em}{1919.4}\\
    (interval, pdf+interval, QDPs, pdf, cdf, none) & & &\\
    \hline
    $\rpos$ text60 & 1170.3 & 851.5 & 1918.7\\
    \hline
    $\rpos$ text85 & 1171.0 & 851.6 & 1918.3\\
    \hline
    $\rpos$ text99 & 1165.0 & 848.1 & 1914.9\\
    \hline
    \end{tabular}
    \vspace{-2mm}
    \caption{The visualization optimal $\rpos$ for different scenarios and visualization conditions.}
    \normalsize
    \vspace{-2mm}
    \label{tab:fernandes-posterior-score}
\end{table}

\item[Rational benchmark:] By taking maximum over visualization optimal, the rational benchmark is the rational agent with full information in \Cref{tab:fernandes-benchmark-score}.
    \begin{table}[htbp]
    \vspace{-4mm}
\small
    \centering
    \begin{tabular}{c|c|c|c}
    \hline
     Scenario ID & 1 & 2 & 3\\
    \hline
    $\rperfect$ & 1171.8 & 852.0 & 1919.4\\
    \hline
    \end{tabular}
    \vspace{-2mm}
    \caption{The rational benchmark $\rperfect$ for different scenarios.}
    \normalsize
    \vspace{-4mm}
    \label{tab:fernandes-benchmark-score}
\end{table}
    
    \item[Value of information:] \Cref{tab:fernandes-infoval} summarizes the value of information $\infoval = \rperfect - \rprior$.
            \begin{table}[htbp]
\small
    \centering
    \begin{tabular}{c|c|c|c}
    \hline
     Scenario ID & 1 & 2 & 3\\
    \hline
    $\infoval$& \multirow{1}{1.8em}{93.1} & \multirow{1}{1.8em}{84.6} & \multirow{1}{1.8em}{69.3}\\
    \hline
    \end{tabular}
    \vspace{-2mm}
    \caption{The value of information $\infoval$ for different scenarios. }
    \normalsize
    \vspace{-4mm}
    \label{tab:fernandes-infoval}
\end{table}
\end{description}


From these calculations, we first note that all visualization conditions have the same visualization optimal, except for the text displays. We quantify this information asymmetry by information loss. 

\begin{description}
    \item [Information loss] We calculate the information loss induced in \Cref{tab:fernandes-info-loss}.
        \begin{table}[thbp]
\small
    \centering
    \begin{tabular}{c|c|c|c}
    \hline
     Scenario ID & 1 & 2 & 3\\
    \hline
   full information & \multirow{2}{0.8em}{$0$} & \multirow{2}{0.8em}{$0$} & \multirow{2}{0.8em}{$0$}\\
    (interval, pdf+interval, QDPs, pdf, cdf, none) & & &\\
    \hline
    text60 & $1.6\%$ & $0.7\%$ & $1.2\%$\\
    \hline
    text85 & $0.9\%$ & $0.6\%$ & $1.6\%$\\
    \hline
    text99 & $7.3\%$ & $4.7\%$ & $6.5\%$\\
    \hline
    \end{tabular}
    \vspace{-2mm}
    \caption{The information loss $(\rperfect-\rpos)/\infoval$ for different scenarios and visualization conditions.}
    \normalsize
    \vspace{-6mm}
    \label{tab:fernandes-info-loss}
\end{table}
\end{description}

All types of visualizations have an information loss $\sim 1\%$, except for text99 which induces a small information loss $\sim 7\%$.

We calculate the cumulative incentive for the rational agent ($\infoval$) across $40$ trials. In the experiment, each $1000$ coins translate into a $\$\convrate$ bonus in real payment, with another $\$1.25$ as a guaranteed base payment, i.e.\ the payment conversion rule is $f(r)=\frac{\convrate}{1000}r + \$1.25$. $d=0.01698, 0.08228, 0.016076$ for scenarios $1,2, 3$, respectively. The value of information for a rational agent in real dollars is shown in \Cref{tab:fernandes-incentive}. Since the information loss for text displays is small ($\leq 7\%$), we omit the payoff calculation for text displays.  

\begin{table}[thbp]
\vspace{-3mm}
\small
    \centering
    \begin{tabular}{c|c|c|c|c}
    \hline
     Scenario ID    & $f(\rprior)$ & $f(\rpos)$ &  $\infoval_f$ & $\infoval_f / f(\rprior)$\\
     \hline
     1    & $\$1.983$ & $\$2.046$ & $\$0.063$ & $3.12\%$\\
     \hline
     2 & $\$3.776$ & $\$4.054$ & $\$0.287$ & $7.37\%$\\
     \hline
     3 & $\$2.440$ & $\$2.484$ & $\$0.044$ & $1.82\%$\\
    \hline
    \end{tabular}
    \caption{$f(\rprior)$ shows the expected payment to a rational agent who takes the optimal fixed action, $f(\rpos)$ shows the expected payment to a rational agent who reads the visualization, while $\infoval_f=f(\rpos)-f(\rprior)$ is the incentive to consult the visualization.}
    \normalsize
    \vspace{-2mm}
    \label{tab:fernandes-incentive}
\end{table}

Across the three scoring rules, the incentive for the rational agent to consult a visualization is always less than $10\%$ of the guaranteed payment of choosing an optimal fixed action (\Cref{tab:fernandes-incentive}). The incentive is not well designed if the goal is to encourage agents to consult the visualizations. 

To improve incentives, we suggest subtracting $f_0$ from all payments, where $f_0$ is a threshold that any behavioral agent's score is unlikely to fall below. For example, one obvious choice of $f_0$ is $30\cdot \reward_0$, obtained by a strategy to always arrive at the bus stop at $30$ minutes. 

Additionally, Fernandes et al.~\cite{fernandes2018uncertainty} conclude from the results of their experiment that with the \text{dot50} visualization, \textit{50\% of decisions will be above 95\% of optimal, about 80\% of decisions will be above 90\% of optimal, and more than 95\% of decisions will be above 80\% of optimal}. However, we find that the baseline is able to achieve a $92.1\%, 90.1\%$, and $96.4\%$ of the optimal for each scenario, respectively, calculated assuming the agent does not look at the visualization. This pre-experimental analysis therefore calls into question how impressive the \text{dot50} performance reported by the original work is, illustrating how without a baseline to compare with, statements based on the proximity of observed behavior to optimal can mislead. 

\subsubsection{Post-experimental Analysis}
\begin{figure}[thbp]
    \centering
    \vspace{-6mm}
    \includegraphics[width = 0.8\columnwidth]{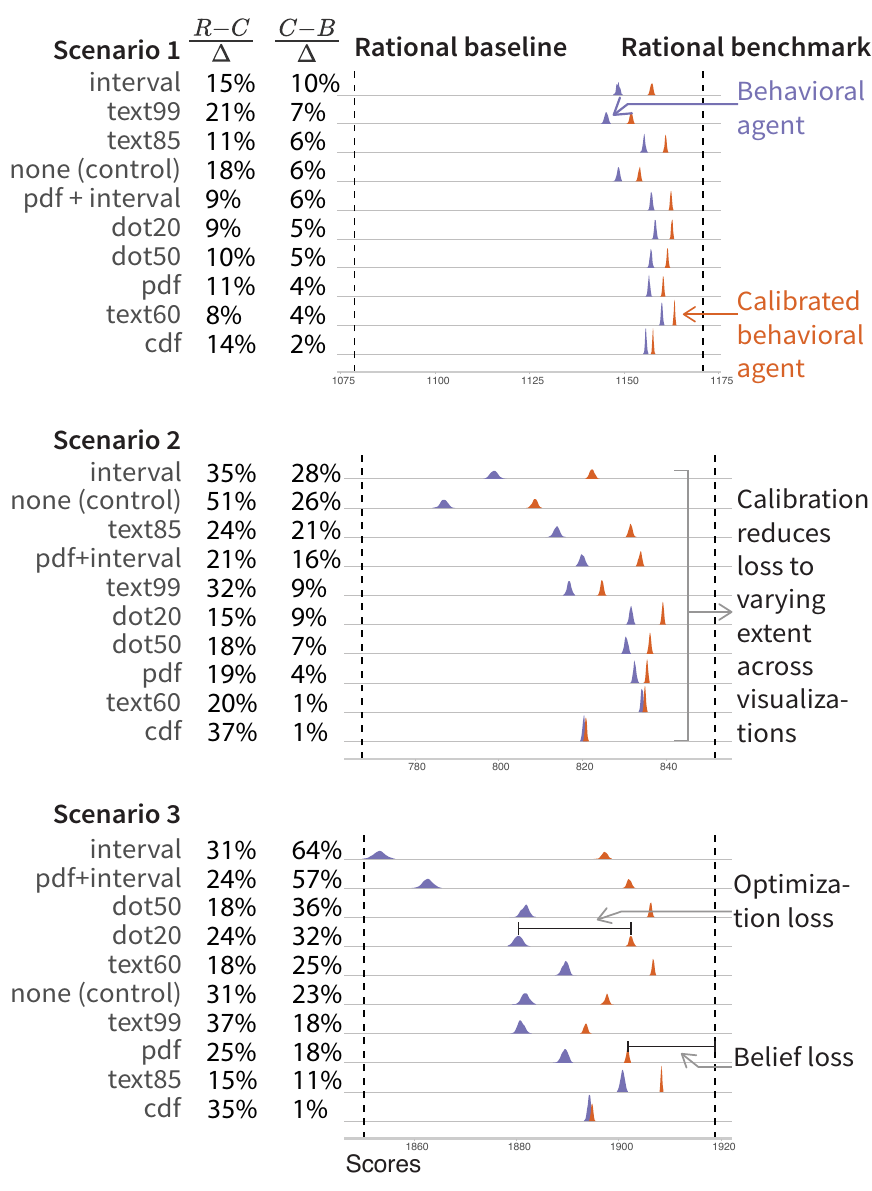}
    \vspace{-6mm}
    \caption{Estimated scores (in simulated coins) for each combination of visualization condition and scenario. Visualizations are ordered by optimization loss for each scenario. The rational agent benchmark $\rpos$ and baseline $\rprior$ are shown as dotted lines,  \textbf{\behavioraltextcolor{behavioral scores}} $\bactionraw$ in purple, and \textbf{\calibratedtextcolor{calibrated behavioral scores}} $\bactioncalib$ in orange.
    }
    \vspace{-6mm}
    \label{fig:fernandes_analysis}
\end{figure}

In our post-experimental analysis, we empirically estimate the behavioral expected payoff $\bactionraw$ for the $10$ visualization conditions in Fernandes et al. The authors fit a mixed-effects Bayesian regression model to predict the ratio \textit{expected/optimal payoff} from visualization condition and trial number, with random effects of scenario and participant. 
Because the outcome ratio is an input to the model, predictions from this model cannot be used to predict expected behavioral scores under different scenarios. We therefore fit our own model to predict agents' actions (i.e.\ chosen arrival time) from visualization condition, scenario, and bus arrival distribution. We include random intercepts by participant and random slopes to allow varying effects of trial number by participant. Full model details and model checks we performed to validate the model are available in supplemental material. 
We use predictions from this model in conjunction with the stated scoring rules in Fernandes et al.\ to calculate expected scores by scenario.\footnote{Even with access to an extended repository containing more complete materials than the public version for the original study, we were not able to exactly reproduce the expected payoffs analyzed by Fernandes et al. However, the expected payoffs our method produces are within 100 simulated coins of their expected payoffs across scenarios.}
Because Fernandes et al.\ did not describe a target distribution of participants over visualization conditions, scenarios, and arrival time distributions, we estimate the behavioral scores by sampling arrival time decisions from our model for the same number of agents they analyzed data from per combination of scenario, visualization condition, and bus arrival distribution. We report scores from $100$ simulated experiments and report the distributions of \textbf{\behavioraltextcolor{behavioral scores}}
(\Cref{fig:fernandes_analysis}, \textbf{\behavioraltextcolor{purple}}). For each simulated experiment, we calculate the \textbf{\calibratedtextcolor{calibrated behavioral scores} }$\bactioncalib$ (\Cref{fig:fernandes_analysis}, \calibratedtextcolor{orange}). In our simulations, we round predicted arrival decisions from our behavioral model to integers to match the format of responses uesd by behavioral agents in the original experiment.

\Cref{fig:fernandes_analysis} shows that behavioral payoffs are above or close to the baseline. Specifically, they are above $\rprior$ for Scenario $1$ and $2$, and above $\rprior$ for Scenario $3$ with the exception of the interval display which induces a payoff below but close to $\rprior$. 

The original paper evaluated visualization conditions in several ways: by plotting estimated learning effects by visualization condition and by ranking visualization conditions by estimated means and standard deviations of the ratio of expected to optimal payoff for the last trial participants completed. All analyses aggregated results across scenarios despite their varying scoring rules. Specifically, ranking visualizations by estimated mean ratio for the last trial resulted in dot50 as the best performing condition, followed by cdf, dot20, text99, text60, pdf-interval, pdf, interval, no uncertainty, and text85. Ranking visualizations by estimated standard deviation of the last trial resulted in similar rankings, with the first portion of the list matching the previous ranking (dot50, cdf, dot20, text99, text60, pdf-interval) but with no uncertainty performing better than pdf and interval in addition to text85.  These rankings lead to the authors' conclusion that dot50 and cdf are the top performing visualizations. 

In contrast, our analysis of behavioral scores shown in \Cref{fig:fernandes_analysis} purple represents expected score over all trials by visualization condition separately by scenario. From these results, dot50 and cdf are not clearly better performing than multiple other visualization conditions (i.e., they are not furthest right in the plot). Ranking by expected behavioral score by scenario leads to text representations as top performing, with text60 ranking best for Scenarios $1$ and $2$ and text85 for Scenario $3$. cdf is ranked sixth, fifth and second while dot50 is ranked fourth, fourth, and eighth for Scenario $1$, $2$, and $3$, respectively. 
These differences compared to the original results may be partially attributable to the different modeling approach (our scores consider expected performance across all trials, not just the last trial) or to slightly differences in our computation of expected ratio compared to theirs, as we were not able to perfectly reproduce their model inputs from the available codebase despite using the equations they provided.
Our ranking of conditions is clearly inconsistent to those of the original paper when it comes to the performance of dot50, which according to Fernandes et al.'s results performed consistently better in expected ratio across the earlier trials as well, with dot50 users starting and ending with higher estimated ratios than any other condition.

\paragraph{Belief Loss} 
The differences between the calibrated score payoff $\bactioncalib$ (\calibratedtextcolor{orange}) and $\rpos$ (rightmost dotted line) show that in general, comparing visualizations by belief loss reduces differences between them compared to raw behavioral scores (\behavioraltextcolor{purple}), and that Scenario $1$ leads to less belief loss than Scenarios $2$ and $3$.
If anything, ranking visualizations by belief loss suggests that 
dot20 performs consistently well (ranked second in all Scenarios). In other words, these visualizations appear to allow users to obtain a good proportion of the available information in the visualization, even if they do not necessarily make the optimal decision from the information.

Visualizations convey over $80\% $ and $61\%$  
of the information to the agents for scenarios $1, 3$, respectively, and over $65\%$ of the information for scenario $2$, with the exception of the no uncertainty control under scenario 2, which conveys $47\%$ of the information ($100\% -$ belief loss $\frac{\rpos - \bactioncalib}{\Delta}$ in \Cref{fig:fernandes_analysis}). 
We conclude that all visualization strategies provide reasonable support for detecting changes in the bus arrival time distributions. Belief loss is not the main source of loss in decision-making. 

\paragraph{Optimization Loss} 
The differences between the calibrated payoff $\bactioncalib$ (\calibratedtextcolor{orange}) and behavioral payoff $\bactionraw$ (\behavioraltextcolor{purple}) suggest that optimization loss is a large source of loss in participants' decision-making.
\Cref{fig:fernandes_analysis} sorts visualization conditions in decreasing order of optimization loss.
We see that interval users have the hardest time optimizing their decisions, while cdf and pdf users are able to do so consistently well (cdf achieving first rank, pdf third rank across Scenarios $1$, $2$, and $3$).
Users of text60 optimize very well except for in Scenario $3$, where their ranking falls from first to sixth. 


%% file: 04_discussion.tex
\section{Discussion}


We contribute rational agent benchmarks for assessing 1) the potential for an experiment to incentivize participants and show differences between visualizations and with best attainable performance, and 2) the sources of error that explain observed results from behavioral agents. As our demonstrations on two celebrated visualization studies show, our framework can be applied to identify improvements in designs and to deepen understanding of results even when the original research was rigorously done. A key feature is that it provides well-defined comparison points for any given visualization, reducing reliance on rough, relative ordering information that is often used to interpret visualization experiment results. 

Returning to the questions posed in \Cref{sec:intro}, by applying our framework we can expect to answer them as follows: 
\begin{itemize}
    \item How hard is the task? The value of information, the difference between rational baseline and benchmark, captures the ``room'' for improvement on the task.
    \item How incentivized are participants? Through pre-experimental analysis, we calculate the expected increase in payment that the participants can get from consulting the visualization.
    \item To what extent do the differences in performance stem from informational asymmetries? This difference is quantified by the information loss.
    \item What are the reasons for sub-optimal decisions from behavioral agents? We separate the sources of loss into 
    \begin{itemize}
        \item the belief loss, the loss from not perceiving the information, and
        \item the optimization loss, the loss from not properly use the information.
    \end{itemize}
    \item To what extent are observed differences driven by ``luck of draw''? Our Bayesian framework compares the expected payoff over the experiment design, avoiding the effect of random lucky draws. 
\end{itemize}
There are many other practical advantages to the rational framework, which we observed in conducting analyses for our demonstrations. For example, having the ability to compare results from different tasks in score space, as we did for Kale et al.~\cite{kale2020visual}, can sidestep the challenges associated with trying to interpret and compare findings between models that estimate different parameters, often under different mathematical transformations that must be inverted to get any perspective on performance from results.
Additional benefits will arise on a case-by-case basis, as demonstrated in our examples.

Integrating measures of the value of information into visualization is an important step forward in the pursuit of more rigorous theoretical foundations for visualization-based inference, as van Wijk called for years ago, and researchers continue to call for today~\cite{dimara2021critical,heine2020towards,hullman2021designing,vanwijk2005value}. By providing a widely applicable definition of a decision task and associated analyses identifying the value of information, our work makes possible deeper connections between information economics and design with data visualization. 
There are many exciting extensions to the rational agent framework to be explored in future work. For example, for certain decisions tasks, such as binary decisions which are amenable to complete characterization, it is likely possible to provide more prescriptive guidelines that can point visualization researchers to the right task to study in the first place given a high-level research goal (e.g., evaluate visualization alternatives for election forecasts).

Another direction worth pursuing is to integrate the rational agent benchmarks into the sample size calculations that experimenters use to ensure that an experiment design is capable of assessing performance differences. We might ask, What sample size is needed to resolve performance with a visualization relative to the value of information to the task? Alternatively, scoring rules could be designed to obtain the same value of information with fewer samples, cf. Li et al.~\cite{li2022optimization} It may also be useful to use quantities from the rational agent framework to contextualize target effect sizes (e.g., in units of $\Delta$) or assumed noise from measurement error (e.g., in units of the standard deviation in scores across trials given the data-generating model) in fake data simulation for power analysis.


\subsection{Limitations}
 Applying the rational agent framework to pre-experiment analysis is not as useful if the experimenter doubts the value of performance incentives, as some have for certain types of behavioral research like crowdsourced experiments (e.g.,~\cite{mason2009financial}). Pre-experiment analysis will not offer actionable guidelines if the experimenter has already predetermined they will provide a flat or no reward scheme. At the same time, choosing to provide no clear incentive to use visualizations in an experiment is usually a signal that the experimenter trusts that their participants will try their best. In such cases, analyzing the value of information is still well-motivated for making sure a study design provides enough room for seeing differences between visualization types and assessing the information gain from any visualization.

The relationship between the rational baseline $\rprior$ and what a participant would do in the actual experiment if they did not look at the visualizations is nuanced. As we describe above, the purpose of $\rprior$ is not to predict how randomizing behavioral agents will score, though in some cases it may. 

The rational agent framework is not intended as a theory of how behavioral agents make decisions. Instead, the benchmarks that the framework provides are valuable in evaluating the quality of decisions of behavioral agents who act differently from a rational one.  While a rational agent would solve such a problem by updating their beliefs based on the empirical joint distribution over signals and states and then choose the optimal action under those beliefs, no intermediate measurement of beliefs is made of the behavioral agent and so his optimization loss cannot be similarly decomposed.  In many experiments, in fact, the behavioral agent is not informed of the prior and, therefore, the Bayesian update is not well defined.  This lack of prior information is also accounted for in the optimization loss.

One of the biggest impediments to applying the framework is not a lack of generalizability but a potential lack of transparent reporting of study details in empirical papers. 
For example, full information about the scoring rule used in a study may not be reported, such as when there are exclusion criteria like performance on an attention check that led to non-payment for a task but not mentioned in the paper. This makes it difficult to analyze the experiment using the rule that the original research used. 

\section{Conclusion}
We contribute a widely applicable analytical framework for benchmarking visualization performance. The approach uses the performance achievable by a rational agent doing the same visualization experiment as a comparison point for the estimated performance of behavioral agents. The framework distinguishes sources of error in results, like not being able to get the information versus not being able to choose the optimal decision given the information one has obtained. Applying the framework to two awarded visualization studies shows how it can identify ways to improve even rigorous decision experiment designs, and enhance the knowledge gained from observed behavioral performance. 

%% file: 05_ack.tex
\section*{Acknowledgment}

We are grateful to Steve Haroz for helpful comments and suggestions that improved this paper. 